\documentclass[prl,twocolumn,showpacs]{revtex4}
\usepackage{amsmath}
\usepackage{amssymb}
\usepackage{amsbsy}
\usepackage{graphicx}
\newcommand{\fett}[1]{{\mbox{\boldmath$#1$}}}
\begin{document}
\title{Translations and Rotations are correlated in Granular Gases}

\author{N. V.  Brilliantov$^{1,2}$, T. P\"{o}schel$^3$, W. T. Kranz$^4$
  and A. Zippelius$^4$}  \affiliation{$^1$Institute of Physics,
  University of Potsdam, Am Neuen Palais 10, 14469 Potsdam, Germany }
\affiliation{$^2$Moscow State University, 119899 Moscow, Russia}
\affiliation{$^3$Charit\'e, Augustenburger Platz 1, 13353 Berlin, Germany} 
\affiliation{$^4$Institute of Theoretical
  Physics, University of Gottingen, Friedrich Hund-Platz 1, 37073 Gottingen,
  Germany}

\date{\today}

\begin{abstract}
  In a granular gas of rough particles the axis of rotation is shown
  to be correlated with the translational velocity of the particles.
  The average relative orientation of angular and linear velocities
  depends on the parameters which characterise the dissipative nature
  of the collision. We derive a simple theory for these correlations
  and validate it with numerical simulations for a wide range of
  coefficients of normal and tangential restitution. The limit of
  smooth spheres is shown to be singular: even an arbitrarily small
  roughness of the particles gives rise to orientational correlations.

\end{abstract}


\maketitle

\begin{figure}[b!]
  \centerline{\includegraphics[width=0.8\columnwidth,angle=0,clip]{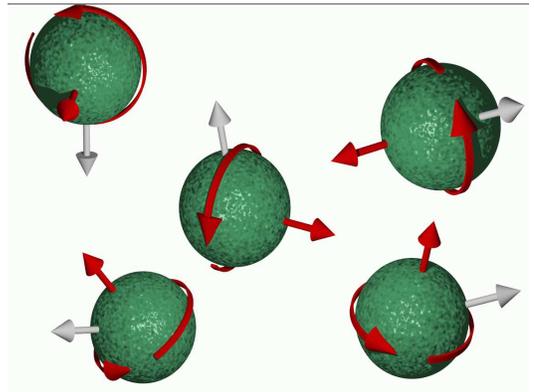}}
\caption{In granular
  gases, depending on the coefficients of restitution, spin (red
  arrow) and linear velocity (grey arrow) are oriented either
  preferably in parallel (cannon ball) or perpendicular (sliced tennis
  ball). This is a fundamental difference to molecular gases, where
  spin and velocity are uncorrelated in orientation.}
  \label{fig:collision}
\end{figure}

Dilute systems of macroscopic particles, called Granular Gases, show
many novel and surprising phenomena when compared to molecular gases.
The particles of Granular Gases are macroscopic bodies which in
general dissipate energy upon collision. As a consequence, such gases
demonstrate features, which drastically differ from molecular gases:
The velocities are {\it not} distributed according to a 
Maxwell-Boltzmann distribution
\cite{GoldshteinShapiro1:1995,NoijeErnst:1998,EsipovPoeschel:1995,BreyCuberoRuizMontero:1999,DeltourBarrat:1997,HuthmannOrzaBrito:2000,Goldhirschetal:2005};
equipartition does {\it not} hold \cite{HuthmannZippelius:1997} and a
homogeneous state is in general {\it unstable}
\cite{GoldhirschZanetti:1993,McNamaraYoung:1992,BritoErnst:1998}.  In
this Letter we present another unexpected result:

The angular and linear velocities of rough particles are correlated in
direction.  In dependence on the coefficients of restitution,
characterizing the dissipative particle properties, the rotational
motion may be preferably perpendicular to the direction of linear
motion, similar to a sliced (spinning) tennis ball, or in parallel to
it, like a rifled bullet, Fig. \ref{fig:collision}. Surprisingly, the
limit of vanishing dissipation of the rotational motion does not
exist, that is, {\em any} arbitrarily small roughness leads to a
macroscopic correlation between spin and velocity. We present a
kinetic theory that quantifies this new effect and find good agreement
with large scale numerical simulations. It is expected that the
reported correlation between spin and linear velocity may have
important consequences in understanding natural Granular Gases, such
as dust clouds or planetary rings.

{\it Model\/}-
We consider a monodisperse granular gas consisting of $N$ hard spheres
of radius $a$, mass $m$ and moment of inertia $I=qma^2$ ($q=2/5$ for
homogeneous spheres addressed here). The degrees of
freedom are the particles' position vectors $\{{\bf r}_i \}$,
translational velocities $\{{\bf v}_i \}$ and rotational velocities
$\{\fett{\omega}_i \}$ for $i=1,2,...N$. The dynamic evolution of the
system is governed by instantaneous two particle collisions, such that
in general {\it both} translational and rotational energy is dissipated.
Introducing the relative velocity at the point of contact
\begin{equation}
{\bf g}={\bf v}_1-{\bf v}_2+a {\bf n} \times ({\fett{\omega}}_1+{\fett{\omega}}_1)
\end{equation}
the collision rules specify the change of $\bf{g}$ in the direction of
${\bf n}=({\bf r}_1-{\bf r}_2)/|{\bf r}_1-{\bf r}_2|$ and
perpendicular to ${\bf n}$:
\begin{equation}
  \label{eq:eps_n_t_def}
({\bf g}\cdot {\bf n})^{\prime} = -\varepsilon_n ({\bf g}\cdot {\bf n})
\qquad
({\bf g}\times {\bf n})^{\prime} = +\varepsilon_t ({\bf g}\times {\bf n}),
\end{equation}
where the primed values refer to the post-collision quantities.  The
coefficients of restitution in normal and tangential direction, $0\le
\varepsilon_n \le 1$ and $-1\le \varepsilon_t \le 1$, characterize the
loss of energy and, thus, describe the slowing down of the linear and
rotational motion of the particles. These coefficients are the central
quantities in the Kinetic Theory of granular
gases \cite{BrilliantovPoeschelOUP}. For $\varepsilon_n=1$ (elastic
spheres) and $\varepsilon_t=\pm 1$ (perfectly smooth/rough spheres)
the energy is conserved, while for $\varepsilon_n \to 0 $ and
$\varepsilon_t \to 0 $ dissipation is maximal.
Combining the collision rules (Eq.
\eqref{eq:eps_n_t_def}) with conservation of angular and linear
momentum, allows one to express the post-collision velocities in terms
of the precollisional ones
\begin{eqnarray} \label{eq:vi_prime}
  {\bf v}_{1}^{\, \prime} & = & 
  {\bf v}_{1} -\fett{\delta} \, ,
  \quad
  {\fett{\omega}}_1^{\, \prime} =
  {\fett{\omega}}_1 +  
   \left( \frac{1}{aq} \right) {\bf n} \times \fett{\delta},\nonumber\\
 {\bf v}_{2}^{\, \prime} & = &
  {\bf v}_{2} + \fett{\delta} \, ,
  \quad
  {\fett{\omega}}_2^{\, \prime} =
  {\fett{\omega}}_2 + \left( \frac{1}{aq} \right) {\bf n} \times \fett{\delta},
\end{eqnarray}
with  ${\fett{\delta}}=
\eta_t {\bf g}+(\eta_n-\eta_t)
({\bf n}\cdot{\bf g}){\bf n}
$ and $\eta_n=(1+\varepsilon_n)/2$, $\eta_t=q(1-\varepsilon_t)/2(1+q)$.

When particles collide according to the above laws, two processes take
place: (i) dissipation of energy and (ii) exchange of energy between
the rotational and translational degrees of freedom. The first process
is described by two time-dependent granular temperatures
\begin{equation}
  \label{eq:Tdef}
  T_{\rm tr}=\frac{m}{3N}\sum_{i=1}^{N} {\bf v}_i^2 \quad
  \mbox{and} \quad
  T_{\rm rot}=\frac{I}{3N}\sum_{i=1}^{N} {\fett{\omega}}_i^2\,,
\end{equation}
The second process drives the system to a quasi
steady-state, which is characterized by a constant ratio of the two
temperatures, $r=T_{\rm rot}/T_{\rm tr}={\rm const.}$ In this state
both temperatures continue to decay with their rates tied together by
$\dot{T}_{rot}/\dot{T}_{tr}=r$. The ratio of temperatures, $r$,
depends on the coefficients of restitution $\varepsilon_n$ and
$\varepsilon_t$
and can take values smaller or larger than one.
If $r<1$ collisions damp the translational motion more efficiently than the rotational one,
$|\dot{T}_{tr}|=|\dot{T}_{rot}|/r >|\dot{T}_{rot}|$,
whereas for $r>1$ the rotations are damped more efficiently.

At first glance there is no reason to expect that linear and angular
velocities ${\bf v}$ and ${\fett{\omega}}$
are correlated. Nevertheless, the exchange of energy
between the rotational and translational degrees of freedom may build
up such correlations. We quantify them by the mean square cosine of
the angle between ${\bf v}$ and ${\fett{\omega}}$:
\begin{equation}
  K\equiv \frac{1}{N}\sum_{i=1}^{N} \frac{ \left(
      {\bf v}_i \cdot {\fett{\omega}}_i \right)^2 }{\left( v_i^2 \omega_i^2 \right)^2} = \frac{1}{N}\sum_{i=1}^{N} \cos^2
  \Theta_i \, . \label{Cdef}
\end{equation}
If there are no correlations between the rotational and translational
motion, as in molecular gases, one obtains $K=1/3$; in granular gases
we find that $K$ in general deviates significantly from $1/3$.

The full quantitative understandig of these correlations requires
detailed mathematical analysis of the collision rules 
Eq. \eqref{eq:vi_prime} (see below). Nevertheless simple physical arguments
are helpful for a qualitative understanding of these correlations. The
transfer of energy from the translational to the rotational degrees of
freedom (and vice versa) depends sensitively on the relative
orientation of ${\bf v}_i-{\bf v}_j$ and ${\fett{\omega}}_i+
{\fett{\omega}}_j$. To sustain a quasi steady state with a fixed
ratio, $r$, of $T_{\rm tr}$ and $T_{\rm rot}$ requires that
fluctuations away from a given $r$ are effectively suppressed. In the
limit of nearly smooth particles, $\varepsilon_t \lesssim 1$, the
argument is particularly simple: In this case, we have $r=T_{\rm
  rot}/T_{\rm tr} \gg 1$ \cite{AspelmeierHuthmannZippelius:2000}, that
is, for most of the collisions the relative velocity of the particles
at the point of contact is dominated by rotations ${\bf g} = {\bf
  v}_i-{\bf v}_j + a {\bf n} \times \left({\fett{\omega}}_i+
  {\fett{\omega}}_j\right) \approx a {\bf n} \times
\left({\fett{\omega}}_i+ {\fett{\omega}}_j\right)$. For nearly smooth
particles the rotational velocities change only slightly upon
collision, ${\fett{\omega}}_i^{\, \prime} \approx {\fett{\omega}}_i$
and ${\fett{\omega}}_j^{\, \prime} \approx {\fett{\omega}}_j$, so that
the collision rule Eq.
\eqref{eq:vi_prime} simplifies to
\begin{equation}
  \label{eq:vi_prime_simple}
  {\bf v}_{i}^{\, \prime}-{\bf v}_{i} \approx -\eta_t {\bf g} \approx
  - \eta_t a\left({\bf n}   \times
    {\fett{\omega}}_i^{\prime} + {\bf n}  
    \times {\fett{\omega}}_j^{\prime} \right) \, ,
\end{equation}
where $\eta_t \sim (1- \varepsilon_t) \ll 1$ . 
The first term on the right-hand side gives a contribution to
${\bf v}_{i}^{\, \prime}$ that is always
perpendicular to the angular velocity ${\fett{\omega}}_i^\prime$. The
second term (of the same order of magnitude), $\eta_t a {\bf n}
\times {\fett{\omega}}_j^\prime$, has no preferred orientation with
respect to $\fett{\omega}_i^\prime$, -- the two vectors
${\fett{\omega}}_i^\prime$ and ${\fett{\omega}}_j^\prime$ being uncorrelated
in a dilute gas. henec, the sum of all contributions leads to
${\bf v}_{i}^\prime$ being preferably perpendicular to
${\fett{\omega}}^\prime_i$.

To describe these phenomena quantitatively beyond the limit of nearly
smooth spheres, we develop an analytical theory which is based on an
ansatz for the $N$-particle distribution function. We assume
homogeneity, except for the excluded volume interaction, and molecular
chaos, implying that the $N$-particle velocity distribution factorizes
into a product of one particle distributions
\begin{equation}
\rho_1({\bf v},{\fett{\omega}},t) \sim
  e^{-\frac{m v^2}{2T_{\rm tr}(t)}}\, e^{-\frac{I
      \omega^2}{2T_{\rm rot}(t)}}\left[1+b(t) v^2
    \omega^2 P_2(\cos\Theta)\right]. \label{eq:ansatz}
\end{equation}
This ansatz takes into account for the first time {\it orientational
  correlations} between linear and angular velocities.  To lowest
order these correlations can be characterized by the second Legendre
polymonial $P_2(x)\equiv\frac{3}{2}x^2-\frac{1}{2}$. In general there
will be higher-order terms in $(\cos \Theta ) $, as well as
non-Gaussian corrections for the velocity and angular velocity
distribution. Here we ignore these terms since the simplest ansatz
already captures, even quantitatively, the correlations of interest.
The ansatz contains three functions, $T_{\rm tr}(t)$, $T_{\rm rot}(t)$
and $b(t)$, which have been calculated self-consistently using the
Pseudo Liouville operator. These functions determine the correlation
of Eq. \eqref{Cdef} according to
\begin{equation}
\langle\cos^2\Theta\rangle=\frac{1}{3}+b(t)\frac{6T_{\rm tr}T_{\rm rot}}{5qm^2a^2}
\end{equation}

\begin{figure}
  \centerline{\includegraphics[width=0.99\columnwidth,clip]{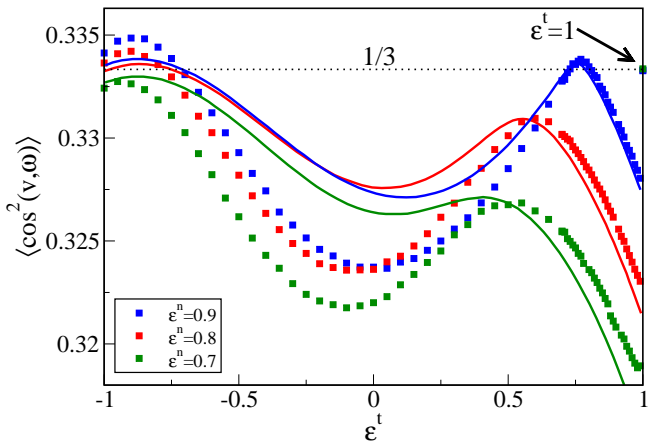}}
    \caption{Steady-state value of the correlation factor $K=\left<
        \cos^2 \Theta_i \right>$ as a function of the coefficient of tangential
      restitution, $\varepsilon_t$, for different values of  normal
      restitution, $\varepsilon_n$.  Lines -- analytical
      theory, points -- DSMC simulations, The isolated point at the
      right border where $\varepsilon_t=1$ indicates vanishing
      correlations for ideally smooth particles.}
      \label{fig:correlations}
\end{figure}

{\it Results\/}-
We find that very strong dynamic correlations may develop, depending
on the initial conditions. After a transient period, the correlation
factor $K$ reaches a steady-state value which is independent of the
initial conditions. In Fig. \ref{fig:correlations} we show this
steady-state value as a function of the coefficient of
tangential restitution $\varepsilon_t$ for several values
of the normal coefficient $\varepsilon_n$. We compare results from
the analytical theory with data obtained with
Direct Simulation Monte Carlo (DSMC)
\cite{Bird:1994} of $N=2\times 10^7$ particles. The analytical results
are in good agreement with simulations,
especially for small dissipation. As can be seen in the figure, the
most pronounced correlations
are observed
for $\varepsilon_t=0$, that is, for maximal damping of the tangential motion.
The correlations weaken when the damping
decreases with increasing $\left|\varepsilon_t\right|$
and are small when the dissipation of the tangential and normal motion
is comparable, i.e. $\varepsilon_n\sim\varepsilon_t$. 
With
still increasing magnitude of $\varepsilon_t$, 
the correlations increase again and $K$
deviates significantly from
the uncorrelated value $1/3$.

\begin{figure}[b!]
  \centerline{\includegraphics[height=0.99\columnwidth,angle=270,bb=120 95 510 758,clip]{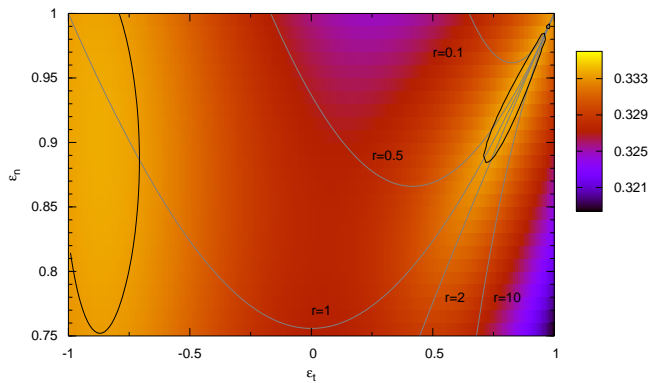}}
    \caption{ Analytical predictions for the steady-state value of
      $K$ as a function of $\varepsilon_n$ and
      $\varepsilon_t$. The black lines show the manifold $K=1/3$ (no
    correlations), the white lines are the contour plots for $r =
    T_{\rm rot}/T_{\rm tr} $.}
      \label{fig:2Dplot}
\end{figure}

The complete dependence of the correlation factor $K$ on both
coefficients of restitution is shown in Fig. \ref{fig:2Dplot} along with
the contour plots of the temperature ratio $r$. In agreement with the
qualitative argument discussed above, we see that translational and
rotational velocities are preferentially prependicular ($K<1/3$) in
those regions where $r$ strongly deviates from $1$.

The right-hand side of Fig. \ref{fig:correlations} reveals that the
limit of smooth spheres, is {\it not continuous} -- contary to naive
expectations. For perfectly smooth particles, $\varepsilon_t=1$, we
obtain $K=1/3$, as expected for a molecular gas. This happens, because
translational and rotational motion decouple, so that all spins simply
persist in time. Hence, trivially, angular and linear velocities are
not correlated for $\varepsilon_t=1$. Nevertheless, even for
vanishingly small roughness $\varepsilon_t \to 1$ the correlation
factor $K$ noticeably deviates from $1/3$. Our analytic theory
predicts
 \begin{equation}
  \label{eq:limit}
  \lim_{\varepsilon_t\to 1}\left<\cos^2\Theta\right> = \frac13 - \frac38\frac{(1 - \varepsilon_n)}{(7 - \varepsilon_n)}\,,
\end{equation}
in good agreement with DSMC.  In the limit
$\varepsilon_t \to 1$ the relaxation time to the quasi steady state
diverges, because the exchange of energy between rotational and
translational degrees of freedom becomes more and more inefficient as their
mutual coupling vanishes.

\begin{figure}
  \includegraphics[width=0.99\columnwidth]{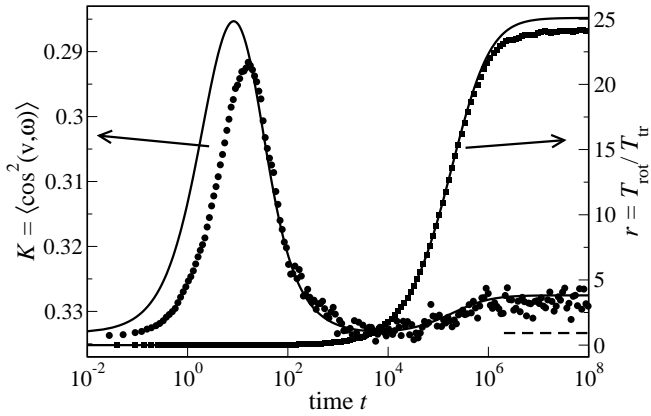}
    \caption{Relaxation of the temperature ratio $r$ and correlation
      factor $K$ to their steady-state values for $ \varepsilon_n=0.8$
      and $\varepsilon_t=0.8$.  The arrows indicate the relevant axes. Full lines: analytical theory, dots:
      simulations of $N=8000$ particles. In the initial state $T_{\rm rot}=0$,
     $T_{\rm tr} =1$, $K=1/3$. Note the three different time regimes:
      First, from $r=0$ to $r=1$, very strong correlations develop;
      second, for $r\sim 1$ correlations are small; finally in the
      quasi steady state with $r\gg 1$, correlations acquire a
      stationary value $K < 1/3$. The dashed line shows
      $K = 1/3$ as
      expected for a molecular gas.}
      \label{fig:dynamics}
\end{figure}

Fig. \ref{fig:dynamics} gives an illustrative example, how
correlations develop in time. We have chosen parameter values and initial
conditions such that both regimes -- dominance of translational motion,
$r\ll 1$ and dominance of rotational motion,  $r\gg 1$ --
are visible. Initially $K=1/3$ and $r \to 0$, which means that correlations
are lacking and the particles' spins are vanishingly small. The 
evolution proceeds in three stages. In the initial stage rotational
motion is generated mainly in grazing collisions so that the particles
rotate around an axis perpendicular to their linear velocity, like a
spinning tennis ball, implying $ \left< \cos^2(\theta) \right> <1/3$.  Once the
rotational motion has become comparable
in magnitude to the translational motion, the correlations are small.
In this intermediate time regime the system is still far from the
quasi stationary state which, for our choice of $\varepsilon_n$ and
$\varepsilon_t$, is characterized by a stationary value of $r\gg 1$.
It is plausible that correlations are small in this intermediate
regime, because both translational and rotational velocities
contribute about equally to the momentum transfer in collision.
Finally in the asymptotic state the
rotational motion is considerably more intense than the translational
motion, so that the quasi stationary state with $r\gg 1$ is
characterized by significant correlations $K<1/3$.

{\it Methods\/}- Two different numerical methods were used: Direct
simulation Monte Carlo (DSMC) \cite{Bird:1994} and Event-driven
Molecular Dynamics (MD) \cite{algo}. In contrast to the latter method,
where the trajectory of each particle of the ensemble is directly
computed according to the basic kinematics and collision rules, the
DSMC is based on the solution of the kinetic Boltzmann equation.
Because the gas is explicitly treated as uniform in this method, and
spatial correlations between grains are ignored, it allows to handle
up to $ \sim 10^8$ particles, which is necessary for the precise
measurement of the effect of interest. While the MD method is more
appealing from a physical point of view, the DSMC is much more
powerful for a homogeneous granular gas; in the limit of low density
the two methods are, in principle, identical \cite{algo}.

In the analytic approach one can use either the Pseudo Liouville
operator technique (for details see
e.g. \cite{AspelmeierHuthmannZippelius:2000}) or equivalently
the Boltzmann equation to derive self-consistent
equations for $T_{tr}(t), T_{rot}(t)$ and $b(t)$. In the quasi
stationary state 
the relative orientation is given by:
\begin{align}
\nonumber
 & \langle\cos^2\Theta\rangle_{\infty}
  - \frac13=\nonumber\\ & - \frac65\frac{A^{(0)} + (A-C)\frac{B^{(0)}}{B}
    + (B^{(0)} + C^{(0)}){r^*}^{-1}}
  {A^{(1)} - 40C + (A-C)\frac{B^{(1)}}{B} + (40B + B^{(1)}){r^*}^{-1}}.\nonumber
\end{align}
Here $r^*$ is the stationary ratio of temperatures which is given
in(\cite{HuthmannZippelius:1997}) together with the coefficients
$A,B$ and $C$. The remaining coefficients are explicitly given by:
\begin{equation}
\begin{aligned}
\nonumber
& A^{(0)} = \frac{16}{3}\frac{\eta_t^3}{q}\left(\frac{2\eta_t}{q} -
1\right)
    - \frac23\frac{\eta_t^2}{q}\left(\frac{8\eta_t}{q} - 3\right)\nonumber\\
   & +  \frac13\frac{\eta_t}{q}\left(\frac{\eta_t}{q} - 1\right)
    + \frac83\frac{\eta_t}{q}\left(\frac{\eta_t}{q} - 1\right)
    \eta_n(\eta_n - 1)\nonumber\\
&    A^{(1)}  =  - \frac{4\eta_t\eta_n^2}{q}\left(\frac{\eta_t}{q} - 1\right) 
    + \frac13\frac{\eta_t^2}{q}\left(\frac{24\eta_t}{q} - 37\right)\nonumber\\
   & -  \frac56\frac{\eta_t}{q}\left(\frac{9\eta_t}{q} - 29\right)
 -\frac{8\eta_t^3}{q}\left(\frac{2\eta_t}{q} - 1\right)
       + \frac43\frac{\eta_t\eta_n}{q}\left(\frac{3\eta_t}{q} - 14\right)\nonumber\\
&    - 12\eta_t\eta_n
    + 22(\eta_t + \eta_n) - 6(\eta_t^2 + \eta_n^2)\nonumber\\
&  B^{(0)}  = \frac13\frac{\eta_t^2}{q}
  \left(\frac{16\eta_t}{q}\left(\frac{\eta_t}{q} - 1\right) + 5\right)\nonumber\\
&    B^{(1)}  =  -\frac23\frac{\eta_t^2}{q}
    \left(\frac{8\eta_t}{q}\left(\frac{\eta_t}{q} - 1\right) + 1\right)\nonumber  
\end{aligned}\notag
\end{equation}
\begin{align}
\nonumber
 &   C^{(0)}  =  \frac23\frac{\eta_t^2}{q}\left(
      8\eta_t(\eta_t - 1) +  4\eta_n(\eta_n - 1) + 3
    \right)\nonumber
  \end{align}
Details of the calculation will be published elsewhere.

{\it Conclusions\/}-
In conclusion, we reveal a novel phenomenon, which is unique for granular
gases and surprising in two respects: a) Except for  very special values of
the coefficients of restitution, $(\varepsilon_n, \,\varepsilon_t)$, the
linear and angular velocities are noticeably correlated. For most of
the parameter values ${\bf v}$ and ${\fett{\omega}}$ are preferably
perpendicular and $K<1/3$ like for a sliced tennis ball.
In a small region of low dissipation they are preferably parallel, so
that $K>1/3$ like for a rifled bullet. b) The limit of vanishing
tangential dissipation, $\varepsilon_t \to \pm 1$ is not continuous.

These results have important consequences for the hydrodynamic theory
of dilute granular flows. It was recently
shown \cite{Goldhirschetal:2005} that the angular velocity needs to be
included in the set of hydrodynamic fields. In view of the singular
nature of the limit of vanishing roughness, perturbation expansions
around the smooth limit are questionable. Orientational correlation
between spin and linear velocity presumably also affect the stability
of the sytem towards shear fluctuations which constitute the dominant
instabilty of granular flows of smooth particles.

{\it Acknowledgments\/}-
This research was supported by a grant from the GIF, the
  German-Israeli Foundation for Scientific Research and Development.


\begin{thebibliography}{15}
\expandafter\ifx\csname natexlab\endcsname\relax\def\natexlab#1{#1}\fi
\expandafter\ifx\csname bibnamefont\endcsname\relax
  \def\bibnamefont#1{#1}\fi
\expandafter\ifx\csname bibfnamefont\endcsname\relax
  \def\bibfnamefont#1{#1}\fi
\expandafter\ifx\csname citenamefont\endcsname\relax
  \def\citenamefont#1{#1}\fi
\expandafter\ifx\csname url\endcsname\relax
  \def\url#1{\texttt{#1}}\fi
\expandafter\ifx\csname urlprefix\endcsname\relax\def\urlprefix{URL }\fi
\providecommand{\bibinfo}[2]{#2}
\providecommand{\eprint}[2][]{\url{#2}}

\bibitem[{\citenamefont{Goldshtein and
  Shapiro}(1995)}]{GoldshteinShapiro1:1995}
\bibinfo{author}{\bibfnamefont{A.}~\bibnamefont{Goldshtein}} \bibnamefont{and}
  \bibinfo{author}{\bibfnamefont{M.}~\bibnamefont{Shapiro}},
  \bibinfo{journal}{J. Fluid Mech.} \textbf{\bibinfo{volume}{282}},
  \bibinfo{pages}{75} (\bibinfo{year}{1995}).

\bibitem[{\citenamefont{van Noije and Ernst}(1998)}]{NoijeErnst:1998}
\bibinfo{author}{\bibfnamefont{T.~P.~C.} \bibnamefont{van Noije}}
  \bibnamefont{and} \bibinfo{author}{\bibfnamefont{M.~H.} \bibnamefont{Ernst}},
  \bibinfo{journal}{Granular Matter} \textbf{\bibinfo{volume}{1}},
  \bibinfo{pages}{57} (\bibinfo{year}{1998}).

\bibitem[{\citenamefont{Esipov and P\"oschel}(1997)}]{EsipovPoeschel:1995}
\bibinfo{author}{\bibfnamefont{S.~E.} \bibnamefont{Esipov}} \bibnamefont{and}
  \bibinfo{author}{\bibfnamefont{T.}~\bibnamefont{P\"oschel}},
  \bibinfo{journal}{J. Stat. Phys.} \textbf{\bibinfo{volume}{86}},
  \bibinfo{pages}{1385} (\bibinfo{year}{1997}).

\bibitem[{\citenamefont{Brey et~al.}(1999)\citenamefont{Brey, Cubero, and
  Ruiz-Montero}}]{BreyCuberoRuizMontero:1999}
\bibinfo{author}{\bibfnamefont{J.~J.} \bibnamefont{Brey}},
  \bibinfo{author}{\bibfnamefont{D.}~\bibnamefont{Cubero}}, \bibnamefont{and}
  \bibinfo{author}{\bibfnamefont{M.~J.} \bibnamefont{Ruiz-Montero}},
  \bibinfo{journal}{Phys. Rev. E} \textbf{\bibinfo{volume}{59}},
  \bibinfo{pages}{1256} (\bibinfo{year}{1999}).

\bibitem[{\citenamefont{Deltour and Barrat}(1997)}]{DeltourBarrat:1997}
\bibinfo{author}{\bibfnamefont{P.}~\bibnamefont{Deltour}} \bibnamefont{and}
  \bibinfo{author}{\bibfnamefont{J.-L.} \bibnamefont{Barrat}},
  \bibinfo{journal}{J. Physique I} \textbf{\bibinfo{volume}{7}},
  \bibinfo{pages}{137} (\bibinfo{year}{1997}).

\bibitem[{\citenamefont{Huthmann et~al.}(2000)\citenamefont{Huthmann, Orza, and
  Brito}}]{HuthmannOrzaBrito:2000}
\bibinfo{author}{\bibfnamefont{M.}~\bibnamefont{Huthmann}},
  \bibinfo{author}{\bibfnamefont{J.}~\bibnamefont{Orza}}, \bibnamefont{and}
  \bibinfo{author}{\bibfnamefont{R.}~\bibnamefont{Brito}},
  \bibinfo{journal}{Granular Matter} \textbf{\bibinfo{volume}{2}},
  \bibinfo{pages}{189} (\bibinfo{year}{2000}).

\bibitem[{\citenamefont{Goldhirsch et~al.}(2005)\citenamefont{Goldhirsch,
  Noskowicz, and Bar-Lev}}]{Goldhirschetal:2005}
\bibinfo{author}{\bibfnamefont{I.}~\bibnamefont{Goldhirsch}},
  \bibinfo{author}{\bibfnamefont{S.~H.} \bibnamefont{Noskowicz}},
  \bibnamefont{and} \bibinfo{author}{\bibfnamefont{O.}~\bibnamefont{Bar-Lev}},
  \bibinfo{journal}{Phys. Rev. Lett.} \textbf{\bibinfo{volume}{95}},
  \bibinfo{pages}{068002} (\bibinfo{year}{2005}).

\bibitem[{\citenamefont{Huthmann and Zippelius}(1997)}]{HuthmannZippelius:1997}
\bibinfo{author}{\bibfnamefont{M.}~\bibnamefont{Huthmann}} \bibnamefont{and}
  \bibinfo{author}{\bibfnamefont{A.}~\bibnamefont{Zippelius}},
  \bibinfo{journal}{Phys. Rev. E} \textbf{\bibinfo{volume}{56}},
  \bibinfo{pages}{R6275} (\bibinfo{year}{1997}).

\bibitem[{\citenamefont{Goldhirsch and Zanetti}(1993)}]{GoldhirschZanetti:1993}
\bibinfo{author}{\bibfnamefont{I.}~\bibnamefont{Goldhirsch}} \bibnamefont{and}
  \bibinfo{author}{\bibfnamefont{G.}~\bibnamefont{Zanetti}},
  \bibinfo{journal}{Phys. Rev. Lett.} \textbf{\bibinfo{volume}{70}},
  \bibinfo{pages}{1619} (\bibinfo{year}{1993}).

\bibitem[{\citenamefont{McNamara and Young}(1992)}]{McNamaraYoung:1992}
\bibinfo{author}{\bibfnamefont{S.}~\bibnamefont{McNamara}} \bibnamefont{and}
  \bibinfo{author}{\bibfnamefont{W.~R.} \bibnamefont{Young}},
  \bibinfo{journal}{Phys. Fluids A} \textbf{\bibinfo{volume}{4}},
  \bibinfo{pages}{496} (\bibinfo{year}{1992}).

\bibitem[{\citenamefont{Brito and Ernst}(1998)}]{BritoErnst:1998}
\bibinfo{author}{\bibfnamefont{R.}~\bibnamefont{Brito}} \bibnamefont{and}
  \bibinfo{author}{\bibfnamefont{M.~H.} \bibnamefont{Ernst}},
  \bibinfo{journal}{Europhys. Lett.} \textbf{\bibinfo{volume}{43}},
  \bibinfo{pages}{497} (\bibinfo{year}{1998}).

\bibitem[{\citenamefont{Brilliantov and
  P{\"o}schel}(2004)}]{BrilliantovPoeschelOUP}
\bibinfo{author}{\bibfnamefont{N.~V.} \bibnamefont{Brilliantov}}
  \bibnamefont{and}
  \bibinfo{author}{\bibfnamefont{T.}~\bibnamefont{P{\"o}schel}},
  \emph{\bibinfo{title}{Kinetic Theory of Granular Gases}}
  (\bibinfo{publisher}{Oxford University Press}, \bibinfo{address}{Oxford},
  \bibinfo{year}{2004}).

\bibitem[{\citenamefont{Aspelmeier et~al.}(2000)\citenamefont{Aspelmeier,
  Huthmann, and Zippelius}}]{AspelmeierHuthmannZippelius:2000}
\bibinfo{author}{\bibfnamefont{T.}~\bibnamefont{Aspelmeier}},
  \bibinfo{author}{\bibfnamefont{M.}~\bibnamefont{Huthmann}}, \bibnamefont{and}
  \bibinfo{author}{\bibfnamefont{A.}~\bibnamefont{Zippelius}}, in
  \emph{\bibinfo{booktitle}{Granular Gases}}, edited by
  \bibinfo{editor}{\bibfnamefont{S.}~\bibnamefont{Luding}} \bibnamefont{and}
  \bibinfo{editor}{\bibfnamefont{T.}~\bibnamefont{P\"oschel}}
  (\bibinfo{publisher}{Springer}, \bibinfo{address}{Berlin},
  \bibinfo{year}{2000}), vol. \bibinfo{volume}{425} of
  \emph{\bibinfo{series}{Lecture Notes in Physics}}, p. \bibinfo{pages}{680}.

\bibitem[{\citenamefont{Bird}(1994)}]{Bird:1994}
\bibinfo{author}{\bibfnamefont{G.~A.} \bibnamefont{Bird}},
  \emph{\bibinfo{title}{Molecular Gas Dynamics and the Direct Simulation of Gas
  Flows}} (\bibinfo{publisher}{Oxford University Press}, \bibinfo{year}{1994}).

\bibitem[{\citenamefont{P\"oschel and Schwager}(2005)}]{algo}
\bibinfo{author}{\bibfnamefont{T.}~\bibnamefont{P\"oschel}} \bibnamefont{and}
  \bibinfo{author}{\bibfnamefont{T.}~\bibnamefont{Schwager}},
  \emph{\bibinfo{title}{Computational Granular Dynamics}}
  (\bibinfo{publisher}{Springer}, \bibinfo{address}{New York},
  \bibinfo{year}{2005}).

\end{thebibliography}
\end{document}